\documentclass[a4paper,12pt]{article}
\usepackage{amssymb,amsthm,latexsym}

\title{A FORGOTTEN PUBLICATION OF ETTORE MAJORANA ON THE IMPROVEMENT OF THE THOMAS-FERMI STATISTICAL MODEL}

\author{Francesco Guerra$^1$ and Nadia Robotti$^2$\\
{\small $^1${\itshape Dipartimento di Fisica, Universit\`a di Roma ``La Sapienza'', and}}
\\
{\small {\itshape INFN, Sezione di Roma1, Piazzale A. Moro 2, 00185 Roma, 
Italy}}\\
{\small e-mail: {\tt francesco.guerra@roma1.infn.it}}\\
{\small $^2${\itshape Dipartimento di Fisica, Universit\`a di Genova}},\\ 
{\small {\itshape via Dodecaneso, 33, 16100 Genova, Italy}}\\
{\small e-mail: {\tt robotti@fisica.unige.it}}
} 
\date{\today}
\begin{document}
\maketitle

\abstract{\noindent Ettore Majorana proposed an important improvement to the Thomas-Fermi statistical model for atoms, with a  communication at the general meeting of the Italian Physical Society, held in Rome, on December 29th, 1928, regularly published on Nuovo Cimento. This communication did not receive any mention, neither in the numerous publications of Fermi and his associates on the subject, nor in the further reconstructions of the life and activity of Ettore Majorana.
The Majorana proposal was not accepted by Enrico Fermi for years, and forgotten. However, it was finally  exploited, without reference, in the 1934 conclusive paper by Fermi and Amaldi on the statistical model for atoms. We analyze the improved model proposed by Majorana, with the help of unpublished notes, kept in the Majorana Archives in Pisa. Moreover, we trace the path leading Fermi to the eventual late acceptance of Majorana improvement, also with the help of the material in  Fermi Archives in Pisa. The conclusion is that Ettore Majorana played an effective leadership in Rome for the very conceptual formulation of the statistical model for atoms.}

\section{Introduction}
On the occasion of the XXIInd General Meeting of the Italian Physical Society, held in Rome at the Physical Institute of the Royal University from December 28th to 30th, 1928, during the session of December 29th, the young Ettore Majorana delivered a communication \cite{Majorana} by the title ``Ricerca di un'espressione generale delle correzioni di Rydberg, valevole per atomi neutri o ionizzati positivamente'' (``Search for a general expression of Rydberg corrections, valid for neutral atoms or positive ions''). This communication is very detailed, and appears in the records of the session, regularly published on the journal of the Italian Physical Society, Nuovo Cimento {\bf VI}, Rivista, pages XIV-XVI, 1929. It was also listed in the German Jahrbuch \"uber die Fortschritte der Mathematik {\bf 55}, 1183 (1929). Now it appears in the electronic archives of Zentralblatt f\"ur Mathematik as follows
\vskip.5cm
\centerline{\bf Zentralblatt--MATH 1931 -- \the\year}
\smallskip
\centerline{\copyright\ European Mathematical Society, FIZ Karlsruhe \& Springer-Verlag 
                  Berlin-Heidelberg}
\vskip1em\hrule\par{\parindent0pt
\vskip1em  55.1183.04 [{\bf 1929}]
\vskip.1em
\par {\bf Majorana, E.}
\par{\it Ricerca di un'espressione generale delle correzioni di Rydberg, valevole per atomi neutri o ionizzati positivamente.}
(Italian)\par
[J] 
Nuovo Cimento 6, XIV-XVI. (1929)\par
\vskip.5em Auszug aus einem Vortrag vor der Societ\`a Italiana di Fisica.
    (Data of JFM: JFM 55.1183.04; Copyright 2004 Jahrbuch Database
    used with permission)\vskip.5em}
\vskip1em\hrule
\vskip.5cm

This communication has  never been referred to in any subsequent publication of Fermi, or his associates, nor in any other publication of the numerous researchers in the field.
Moreover, in the first very authoritative biography of Ettore Majorana, published in 1966 by Edoardo Amaldi \cite{Amaldi}, \cite{Amaldi1}, the communication does not appear in the list of published scientific works. As a consequence, by a natural dragging effect, it has been never mentioned in any further published work on the life and scientific activity of Ettore Majorana. We refer for example to the recent extensive account given in \cite{Bonolis}, where all references to the previous literature on the subject can be found, together with a deep comprehensive description of the cultural and scientific athmosphere surrounding Majorana activity. 

As we will show, Enrico Fermi did not accept for many years the proposal made by Majorana in 1928. Surely he knew it quite well, because he was present at the Meeting, and gave also two communications in the same session, one before Majorana, and one after. Moreover, at that time Majorana was still a student, closely associated to Fermi.  In 1929 Majorana moved  his activity toward nuclear physics, and earned his doctoral degree in Rome, on July 6th, 1929, by presenting a research thesis on the ``Quantum Theory of Radioactive Nuclei'', under Fermi supervision. 

The basic formulation of the statistical model (see \cite{Fermi}, and papers n. 44, 45, 46, 47, 48, 49, in Fermi Collected Papers \cite{collected}) was not changed by Fermi, and was extended to positive ions along the same lines in 1930 \cite{Fermiioni}, without taking into account Majorana suggestions. Moreover, the extensive research on the applications of the Fermi statistical model, done by Fermi himself and his associates in Rome up to 1933-34, always exploited the original Fermi statistical formulation, with the 1930 extension for positive ions.

In 1934, a monumental work by Fermi and Amaldi \cite{FermiAmaldi} appeared in the Proceedings of the Royal Academy of Italy, communicated in the session of May 18th, 1934. This paper has to be considered as the final culminating work on the applications of Fermi statistical model done in Rome. It appeared in a period of intense experimental activity on nuclear physics by Fermi and his associates. In fact, Fermi had announced the discovery of neutron induced artificial radioactivity \cite{AGR} on March 25th, 1934.  After that, Fermi and his associates in Rome were deeply involved in the systematic exploration af the activation properties of all known elements along the periodic table. In particular, the very important paper \cite{transuranic} (n. 86a-86b in \cite{collected}), signed by all members of the research group and announcing the discovery of transuranic elements, is dated May 10th, 1934.

The Fermi-Amaldi paper is based, as enphasized by the authors, on an ``improved'' statistical model. The main improvement, concerning the effective potential acting on the optical electron, is exactly what was proposed by Majorana, more than five years before.

The purpose of our paper is to give a detailed description of Majorana contribution in \cite{Majorana} to the improvement of the formulation of the Thomas-Fermi statistical model for atoms, in the frame of the proper conceptual and historical perspective. To this purpose, we will rely also on unpublished notes, kept in the Majorana Archives at the Domus Galilaeana, in Pisa. For an account of the material in the Archives, see the catalogue in \cite{Amaldi}.
 
We will try also to trace the reasons that led Fermi to eventually accept Majorana proposal. To this purpose, we will rely not only on the published literature, but also on the unpublished notes kept in the Fermi Archives \cite{LRS}, also at the Domus Galileaena in Pisa.

The conclusion of this paper is that Ettore Majorana played an effective leadership in Rome for the very conceptual formulation of the statistical model for atoms.

This paper is organized as follows. In Section 2, we recall the basic structure of the Thomas-Fermi statistical model \cite{Thomas}, \cite{Fermi}, and describe Fermi program for the exploitation of the model. Then we mention the first important results obtained in 1928 by Fermi himself, by Franco Rasetti \cite{Rasetti}, and by Giovanni Gentile jr with Ettore Majorana \cite{GentileMajorana}. Then, in Section 3, we describe Majorana proposal of improvement, as published in the mentioned communication \cite{Majorana}, also with the help of the Pisa unpublished Majorana notes. 
Finally, in Section 4, we will try to describe the path followed by Fermi, leading to the eventual final acceptance of Majorana proposal. 

Section 5 is devoted to some conclusions and outlook for further developments along the program of reconstructing the true scientific personality of Ettore Majorana, beyond the composite legends, that have been developed during the years, after his misterious disappearence in March 1938.
On the eve of the centennial of his birth (August 5th, 1906) we consider this task as a real duty, for all people interested in the history of ideas.

\section{Fermi statistical model for the atom}

During the session of December 4th, 1927, at the Accademia dei Lincei, in Rome, Orso Mario Corbino, member of the Academy and also director of the Royal Institute of Physics of the University of Rome, presented a note \cite{Fermi} by Enrico Fermi, with the title ``Un metodo statistico per la determinazione di alcune propriet\`a dell'atomo,'' (``A statistical method for the determination of some properties of the atom''). Fermi ideas are very simple and very powerful. In principle, according to quantum mechanics, an atom should be described by the full Schr\"odinger equation, which is clearly untractable when the number of electrons becomes large. Fermi treatment is based instead on a kind of electrostatic effective mean field, produced by the nucleus and all electrons. The electrons are considered as a completely degenerate ensemble, obeying Fermi statistics, under the influence of the previous electrostatic potential. Fermi statistical distribution  gives the electron density as a function of the potential. On the other hand, the potential satisfies the Poisson equation, with sources given by the nucleus and the electron mean distribution. This nonlinear problem is easily solved by relying on the numerical integration of a second order differential equation, with appropriate boundary conditions. Fermi treatment was originally presented for an atom at any temperature, but clearly only the zero temperature case is relevant for atoms in the usual conditions.

Fermi was unaware that an essentially equivalent scheme had been presented by Llewellen Hilleth Thomas  in the paper \cite{Thomas}, sent to the Cambridge Philosophical Society on November 6th and read on November 22nd, 1926. However, Fermi program was much more ambitious, aiming at a systematic exploitation of atomic properties by using the model. In fact, already during the first half of the year 1928 a series of papers appeared, where the following atomic problems were considered. Fermi himself made applications to the periodic system of elements, by locating the anomalous groups, and evaluated Rydberg correction for the spectroscopic 
$s$-terms. Then Franco Rasetti \cite{Rasetti} calculated the $M_3$ R\"ontgen terms.
Moreover, Giovanni Gentile jr and Ettore Majorana, in the joint work \cite{GentileMajorana},  calculated the energy and the level splitting for the $3d$ R\"ontgen term of Gadolinium $(Z=64)$, and Uranium $(Z=92)$, and for the $6p$ optical term of Caesium $(Z=55)$. Moreover, they calculated also the intensity ratio for the first two absorption lines of Caesium. All results, obtained by Fermi and his associates in the first half of 1928, were summarized by Fermi in a very detailed review \cite{FermiFalkenhagen} (n. 49 in \cite{collected}), based on his report at the conference Leipziger Tagung, held in Leipzig on June 1928.

Let us recall, with a minimum of technical details, the main features of the Fermi statistical model, and its applications.

Call $n$ the electron density and $V$ the electric potential at each point. According to Fermi statistical distribution for a completely degenerate electron gas, the density and the potential are connected through 
\begin{equation}\label{fermidist}
n=\frac{2^{9/2}\pi m^{3/2} e^{3/2}}{3 h^3} V^{3/2},
\end{equation}
where $m$ and $e$ are the electron mass and charge, and $h$ is Planck's constant. In this expression Fermi has also inserted an additional statistical weigth $2$, in order to take into account the two possible orientations of the spinning electron.

The electric density at each point is given by $\rho=-en$. Therefore the electrostatic Poisson equation gives
\begin{equation}\label{poisson}
\Delta V = 4\pi n e = \frac{2^{13/2}\pi^2 m^{3/2} e^{5/2}}{3 h^3} V^{3/2},
\end{equation}
where $\Delta$ is the Laplace operator.

Due to the obvious rotational symmetry, all involved functions do depend only on the distance $r$ from the nucleus. Moreover, by calling $Z$ the atomic number of the atom, the following two boundary conditions must be satisfied 
\begin{eqnarray}\label{origine}
\lim_{r\to 0} r V(r) & = Z e,\\ \label{carica}
\int n d\tau & = Z,
\end{eqnarray}
where $d\tau$ is the volume element.
The first says that near the nucleus the potential must have the usual Coulomb form under the influence of the charge of the nucleus, since the effect of all electrons is completely screened. The second tells us that the total number of electrons must be $Z$.

Through the definitions
\begin{equation}\label{ansatz}
x=r/\mu,\ \ \ \phi=x V / \gamma,
\end{equation}
where
\begin{equation}\label{mu}
\mu=\frac{3^{2/3} h^2}{2^{13/3}\pi^{4/3} m e^{2}Z^{1/3}},\ \ \ \gamma=\frac{2^{13/3}\pi^{4/3} m Z^{4/3}e^{3}}{3^{2/3} h^2}, 
\end{equation}
by using (\ref{fermidist}) and (\ref{poisson}), we arrive to the following celebrated Fermi equation for $\phi$
\begin{equation}\label{fermieq}
\phi^{\prime\prime}=\phi^{3/2}/\sqrt{x},
\end{equation} 
where $\prime\prime$ denotes the second derivative with respect to $x$.

The boundary conditions (\ref{origine}) and (\ref{carica}), give boundary conditions for $\phi$, in the form
\begin{equation}\label{bc}
\phi(0)=1,\ \ \ \phi(\infty)=0.
\end{equation}

The physical interpretation of the Fermi function is very simple. In fact, the potential has the following expression
\begin{equation}\label{pot}
V=\frac{Ze}{r}\phi(\frac{r}{\mu}).
\end{equation}  
Therefore, the function $\phi$ describes the screening effect of the electrons on the Coulomb potential of the nucleus.

This scheme is exploited by Fermi in order to determine the energy of any electronic level. Consider an atom with atomic number $Z$, and one of its electrons. All the other $Z-1$ electrons are treated in the frame of the statistical model. To the first approximation, Fermi choice of the effective potential energy for the selected electron, by the combined effect of the nucleus and the other electrons, is the following
\begin{equation}\label{effpot}
U(r)=- \frac{e^2}{r}(1+(Z-1)\phi(\frac{r}{\mu})),
\end{equation}
where for $\mu$ one has to exploit the value obtained from formula (\ref{mu}) putting $Z-1$ in place of $Z$.

The effective potential is put into the Schr\"odinger equation for the electron, in order to determine its energy levels.

The scheme is very simple and powerful. Simple calculations give numerical results in reasonable agreement with the experimental findings.

Fermi extended this scheme to the case of positive ions, with ionization $z$, in his paper \cite{Fermiioni}. The modifications are the following. There is a finite radius $x_0$ for the electronic distribution. Fermi function $\phi$ is found to satisfy the usual equation (\ref{fermieq}), for $x<x_0$, while for $x>x_0$ we have $\phi^{\prime\prime}=0$. Now the boundary conditions are
\begin{equation}\label{bcion}
\phi(0)=1,\ \ \ \phi(x_0)=0,\ \ \ -x_0 \phi^{\prime}(x_0)=\frac{z}{Z}.
\end{equation}
In the case of ions, the potential turns out to be
\begin{equation}\label{potion}
V=\frac{Ze}{r}\phi(\frac{r}{\mu})+ \frac{ze}{\mu x_0},
\end{equation}
while the effective potential energy for a single electron is chosen as
\begin{equation}\label{effpotion}
U(r)=- \frac{e^2}{r}(1+(Z-1)\phi(\frac{r}{\mu}))-\frac{z e^2}{\mu x_0}.
\end{equation}
Fermi scheme for positive ions reduces to the previous one for neutral atoms if $z=0$.
   
The Thomas-Fermi model, even though it gives only a very crude approximation to real atoms, met an extraordinary success. During the years, hundreds and hundreds of papers have been published on the subject, dealing with various improvements and applications, in different fields. For comprehensive reviews on the physical applications we refer for example to \cite{Gombas}, \cite{March}, \cite{Spruch}. Even from a mathematical point of view, the model attracted the attention of many researchers, starting from 
the very early 1934 work of Carlo Miranda \cite{Miranda}, and previous references quoted there. A nice review on mathematical results has been given in 1981 by Elliot Lieb \cite{Lieb}. For recent results see \cite{Siedentop}. 

The mathematical interest, as Lieb shows, is mainly connected to the fact that the ground state energy of the Thomas-Fermi atom is the same of the true Schr\"odinger atom, in the leading term, for high values of the atomic number $Z$. In fact, for the true ground state energy $E$ of the related Schr\"odinger equation we have  
\begin{equation}\label{energy}
E=-c_{TF} Z^{7/3}+ O(Z^2),
\end{equation}
where the leading term is exactly the Thomas-Fermi expression, and the constant $c_{TF}$ can be explicitely calculated, as already shown by Fermi in \cite{FermiFalkenhagen}.
Next $O(Z^2)$ correction is very delicate, and it has been rigorously established only in relatively recent times \cite{Siedentop}.

\section{The forgotten publication. An improvement to the Thomas-Fermi model}
               
The involvement of Ettore Majorana with Fermi statistical model started very early. At the beginning of 1928, Majorana, as a student at the University of Rome, moved from the School of Engineering, where he had enrolled in the Fall of 1923, to the Faculty of Sciences, in order to continue his studies in Physics. 

In few months, he acquired a very deep knowledge of the structure of Fermi statistical model. In fact, in his notebook ``Volume II'', kept at the Domus Galilaeana in Pisa, reporting at the beginning the date of April 23rd, 1928, we can find, among other things, a very clever calculation of the values of the Fermi function in section 8, an evaluation of the infra-atomic potential without statistics in section 9, some applications of Fermi potential in section 10, and the statistical curve of the fundamental terms in neutral atoms in section 11. 

Along the program of applications of the statistical model, at the beginning of 1928, Ettore Majorana began a fruitful collaboration with  Giovanni Gentile jr., then in Rome. Their first results were published in the already mentioned joint paper \cite{GentileMajorana}. An early draft of this paper is kept in the Majorana archives in Pisa. Handwritten parts produced by the two coworkers are easily recognizable. The part written by Majorana is extremely interesting, because it contains some deep considerations about the limits of the statistical model, which have not been inserted in the printed version.  The scientific collaboration between the two young researchers continued after this joint paper, on some other topics, as the existing correspondence shows. However, they preferred to publish separately their further results on atomic physics, may be because joint papers were considered less valuable in academic competitions.

In his December 1928 communication to the Italian Physical Society, published on Nuovo Cimento, Majorana presented  results about his improvement of the Fermi model, and its extension to positive ions. The paper of course is in Italian, and is written in the form of a record of the session, but is very clear and detailed. Here we present a translation of the first part, where the essence of the improvement is presented, and a description of the  content following the translated part.

\noindent
Translation from Nuovo Cimento {\bf 6}, page XIV (1929).

Majorana dr Ettore: \textit{Search for a general expression of Rydberg corrections, valid for neutral atoms or positive ions}.

\noindent It is an application, the Author says, of the statistical method devised by Fermi. In the interior of an atom with number $Z$, $n$ times jonized, the potential can be put in the form
$$
V=\frac{Ze}{r}\phi(x)+ C,
$$
where $x$ is the distance measured in units
$$
\mu=0.47 Z^{-\frac{1}{3}}(\frac{Z-n}{Z-n-1})^{\frac{2}{3}} 10^{-8} cm,
$$
$\phi$ obeys to a well known differential equation and to the boundary conditions
$$
\phi(0)=1,\ \ -x_0 \phi^{\prime}(x_0)=\frac{n+1}{Z}\ \ \ {\textstyle beeing}\ \ \ \phi(x_0)=0,
$$
and $C$, which is the potential at the boundary of the ion, has the value
$$
C=\frac{(n+1) e}{\mu x_0}.
$$

In the previous formula we do not consider the local potential, but the mean effective potential acting on some electron in any given given point in space. The two potentials, which to the first approximation are identical, are to be kept distinct, as we have now tacitly understood, in the second approximation, in order to take into account that the elementary charge of an electron is not vanishing, but has a finite value. As a matter of fact, we can not proceed to the second approximation in a rigorous way, but, in the case of an isolated atom, one can imagine some quite satisfactory methods. Among these, the simplest one leads to the expression mentioned before.

\noindent (end of the translation of the first part of Majorana communication)

The communication continues with a very elegant expression of the Rydberg corrections to the spectral lines, in closed form. In the second part of the communication, Majorana gives a preliminary account about an attempt of statistical evaluation of the effect of chemical bonds on the R\"ontgen spectral lines. He considers the elements Al, Si, Ph, S, and their molecular compounds with Oxygen. The displacement of the spectral lines in the compounds is interpreted in terms of the simple elements with some effective ionization, by using the previous theory. This extremely interesting line of research does not appear to have produced further published results.

Further clarification of the improvement proposal put forward by Majorana, can be found in the notebook ``Volume II'' in Pisa, in section 16. Here also we give a complete translation of this short section.

\noindent Domus Galilaeana (Pisa). Majorana Archive. Notebook ``Volume~II'', section 16.

16 - Potential inside the atom in 2nd approx.

\noindent
From the statistical relation between the effective potential and the density
$$
\rho = \kappa (V-C)^{\frac{3}{2}}\ \ \ \ \ \ \ \ \ \ \ \ \ \ \ \ \ (1)
$$
from the Poisson equation satisfied by the local potential
$$
\Delta_2 V_0 = -4\pi\rho\ \ \ \ \ \ \ \ \ \ \ \ \ \ \ \ \ \ (2)
$$
and from the approximatively verified relation, for an $n$ times ionized atom
$$
\Delta_2 V = \frac{Z-n-1}{Z-n} \Delta_2 V_0\ \ \ \ \ \ (3)
$$
one gets through the elimination of (2):
$$
\Delta_2 V = -4\pi\rho \frac{Z-n-1}{Z-n}\ \ \ \ \ \ (4)
$$
the potential in the interior of the ion:
$$
V = \frac{Ze}{r}\phi(\frac{r}{\mu})+C
$$
beeing: $\mu=0.47 Z^{-\frac{1}{3}}(\frac{Z-n}{Z-n-1})^{\frac{2}{3}}$ \"A
$$
\phi^{\prime\prime}=\frac{\phi^{\frac{3}{2}}}{\sqrt x} \phi(0)=1,  -x_0 \phi^{\prime}(x_0)=\frac{n+1}{Z}, \phi(x_0)=0
$$
$$
C=\frac{(n+1) e}{\mu x_0}
$$
(end of Majorana note)

The very essence of Majorana improvement can be easily recognized. In Fermi, the statistical distribution of electrons is ruled by the microscopic field, while in Majorana it is ruled by the effective field, and a Lagrange multiplier is added, even for neutral atoms, which shifts the value of the Fermi energy surface. As a result, for Majorana, even neutral atoms have a finite radius. From a physical point of view, in Fermi atoms each electron is interacting also with itself. In Majorana improvement this self-interaction is avoided, through a very simple average argument, giving the relation between $V$ and $V_0$.

After the December 1928 communication, Majorana, strangely enough, did not publish any paper on the subject. At the beginning of 1929, the interests of the young Majorana moved toward nuclear physics. In fact, he earns his doctorate in Physics on July 6th, 1929, with a research thesis on quantum mechanics of radioactive nuclei. 

\section{Early refusal and late eventual acceptance of the Majorana improvement}

Fermi did not accept  Majorana proposal for many years. In fact, all subsequent work in Rome, done under influence of Fermi, up to the year 1933, exploits the effective potential in the original Fermi scheme. As a significant very late example, this can be clearly seen by looking at the joint 1933 Fermi-Segr\`e paper \cite{FermiSegre}, devoted to the relativistic evaluation of the hyperfine structure effects on atomic spectra, due to nuclear magnetic momenta. Here, in section 2, formula (14), Fermi and Segr\`e still exploit the original Fermi form (\ref{effpot}) for the effective potential.

Moreover, in the session of March 21st, 1930-VIII, of the Royal Academy of Italy, Fermi presented his previously mentioned theory of the statistical model for positive ions \cite{Fermiioni}, in a form coherent with his version for neutral atoms, and therefore different from Majorana December 1928 improvement. This scheme was further exploited by his associates, see for example the 1930 contribution of Segr\`e \cite{Segre}. 

All results, obtained until 1933, by Fermi and his associates, where summarized in tables for the wave functions of different atomic levels, in the notebook n. 18, called ``Thesaurus $\psi$-arum'', kept in the Fermi Archives in Pisa \cite{LRS}. This notebook contains information about the spectroscopic levels of the elements Cd$_{48}$, Sn$_{50}$, V$_{23}$, Pb$_{82}$, Tl$_{81}$, W$_{74}$.

During the session of May 18th, 1934, of the Royal Academy of Italy, Fermi presented the final conclusive report \cite{FermiAmaldi} about all work done by himself and his associates on the subject of the study of atomic spectra using the statistical model. This monumental paper, after a theoretical introduction, contains tables for the wave functions $\infty$s, \textit{i.e.} for zero angular momentum and zero energy, of the following 14 elements Ne$_{10}$, Si$_{14}$, K$_{19}$, Fe$_{26}$, Ga$_{31}$, Rb$_{37}$, Mo$_{42}$, Ag$_{47}$, I$_{53}$, Ce$_{58}$, Ho$_{67}$, W$_{74}$, Hg$_{80}$, U$_{92}$. The basic structure of the statistical model allows to infer properties for other elements, and other wave functions, through a smooth interpolation. The paper comes very late, in fact it is already mentioned in the previous Fermi-Segr\`e paper \cite{FermiSegre}  on hyperfine structures, of one year before. Moreover, it is very surprising that Fermi and Amaldi invest a so great effort, of writing and editing, exactly during the months where they were very actively involved in the exploration of the periodic table in search of neutron induced radio-activity, after Fermi discovery announced on March 25th, 1934.

The Fermi-Amaldi paper is based on an ``improved'' statistical model, as declared by the authors. The improvement goes along two lines. In the first place there is a change in the general formulation, exactly along the line of Majorana proposal of five years before. It is amusing to check that the initial theoretical section 2, devoted to the calculation of the effective potential, in \cite{FermiAmaldi} (n. 82, page 594 in \cite{collected}), repeats, almost \textit{verbatim}, the content of Majorana communication and notes, with a mild change of notations. Surely it is a triumph for Majorana ideas. However, in 1934 Ettore had already secluded himself, apparently withdrawing from active research. 

We do not know his reactions to this further substantial success of his own, even in absence of citation and recognition. The year before, when he received a copy of the Fermi-Segr\'e paper \cite{FermiSegre}, in a letter \cite{BGentile} to Giovanni Gentile jr., dated Copenhagen, March 12th, 1933, he made the following comment: ``Ho avuto da Roma una copia della grande opera di Fermi e Segr\`e che apparir\`a presto fra le memorie dell'Accademia. A questa dovr\`a seguire un'altra grande opera di Fermi e Amaldi sui calcoli statistici.'' (``I have received from Rome a copy of the great work of Fermi and Segr\`e which will appear soon among the Memoirs of the Accademy. To this it will follow another great work of Fermi and Amaldi on the statistical calculations.'')    

The other improvement in the Fermi-Amaldi paper is the exploitation of a relativistic form for the wave equation. In the Fermi-Segr\`e paper of 1933 \cite{FermiSegre} (n. 75b in \cite{collected}), at the end of Appendix I, where relativistic wave functions were exploited for the calculation of the hyperfine effects, Fermi and Segr\`e say ``Ringraziamo il dott. E. Majorana per varie discussioni relative ai calcoli di questa Appendice'' (``We thank dr. E. Majorana for  various discussions related to the calculations in this Appendix'', page 533 of \cite{collected}).  Therefore, surely Majorana was involved in the exploitation of relativistic wave equations in the frame of the statistical model. His research notebooks in Pisa afford a full confirmation of this involvement. In conclusion, it seems that Majorana played some role also for this second improvement. 

Here, we note also a kind of parallelism between the involvement of Majorana with the statistical theory of the atom, and his involvment with the structure of the exchange forces in the nucleus. It is very well known that Majorana, during his visit in Lipsia, starting in January 1933, found a very relevant way to improve the formulation of the nuclear model developed by Heisenberg, by changing the form of the exchange forces \cite{MajoranaKern}. Heisenberg immediately accepted the improvement made by Majorana to his theory, and gave a large credit to his young guest in Leipzig, for example on the occasion of the  1933 Solvay meeting \cite{Heisenberg}. In this way the contribution of Majorana was immediately recognized. While, in the previous analogous case of the improvement of the statistical model, acceptance came very late, when even recognition and priority had become irrelevant for the still young Majorana, already apparently out of active research.

It is also important to analyze the reasons for which Fermi eventually accepted Majorana proposal. Surely his statistical model was subject to criticism. Let us give an example. 

In the 1932 paper \cite{Braunbeck} by  
Werner Braunbeck,
some criticism is voiced, on physical ground, against the picture provided by the Thomas-Fermi theory for atoms and ions, especially as far as their effective radius is concerned. In particular, it is remarked that the Thomas-Fermi theory produces a very important difference between the neutral atoms and the positive ions, since it gives to atoms an infinitely extended charge distribution and to ions a finite one. This difference surely does not rely on the very nature of atoms and ions, but on the contrary comes in through the method as a foreign feature. Moreover, it is recalled that the positive ion can not be treated in the frame of the Thomas-Fermi theory, as it has been also recognized in the rigorous treatment, described in \cite{Lieb}.

These defects of the original Thomas-Fermi formulation of the statistical model, can be simply understood, by taking into account that here each single electron is interacting also with itself, because the repulsion is described through the overall electric density.

On the other hand, Majorana improvement tries to correct these drawbacks, by the simple device of letting the effective potential, and not the microscopic potential,  influence the electronic Fermi distribution, and by introducing the related Lagrange multiplier modifying the Fermi surface, as it is shown by the Majorana formula (2), in his notebook,  as compared with the Fermi's (\ref{fermidist}).

As a simple consequence, Majorana neutral atom has a finite radius.

Moreover, it is very simple to realize that Majorana improvement allows for stable positive ions of charge one (in this case with an electronic density extending to infinity).

Fermi shift toward Majorana version of the model is registered in a very impressive way in the mentioned notebook n. 18 in Pisa, called ``Thesaurus $\psi$-arum''. The notebook is exploited also on the reverse side, for theoretical calculations, according to a well known Fermi practice (see for example \cite{AGR}). In the first pages we find various calculations exploiting the old formulation. In particular, the solution for the variations to the Thomas-Fermi equation is given in closed terms, with a detailed table of the calculated values. Some details follow on the calculation of the statistical distribution of electrons in a ion, and of the associated wave function, with extended tables. By looking at the written form of the effective potential, we can easily recognize that Fermi was exploiting the old formulation.  Then suddenly, at page 10$^*$, the essential ingredients of the improved formulation are introduced, without any explanation. We immediately recognize, in clean Fermi handwriting, the new forms, already presented by Majorana many years before, for the rescaling factor $\mu$  (in page 6$^*$ $\mu$ still appears in the old form  (\ref{mu})), for the boundary conditions on $\phi$, and for the effective potential, while it appears in the old form at page 4$^*$.

The change in Fermi evaluation of the proposed Majorana improvement should have been developed around the middle of 1933, since the old effective potential still appears in the Fermi-Segr\`e paper, presented on March 10th, 1933. This change prompted the correction of all calculations done so far in the announced Fermi-Amaldi paper. This is a good possible explanation for the late appearance of the final version \cite{FermiAmaldi}.

\section{Conclusion and outlook for future research}  

So far, the known recognized contributions of Ettore Majorana to the Thomas-Fermi statistical model for atoms were restricted to his 1928 joint paper with Giovanni Gentile jr \cite{GentileMajorana}, and to his clever method of solving the Thomas-Fermi equation, as reported in section 8 of the Majorana unpublished notebook, called ``Volume II'', in the  Archives in Pisa (see also \cite{Esposito}). However, we have shown that Majorana has given also an important contribution to the conceptual improvement of the statistical model, clearly expressed in his communication to the meeting of the Italian Physical Society on December 29th, 1929, and further clarified in section 16 of his notebook ``Volume II''.

Let us explicitely notice that our reconstruction is mostly based on the 1928  Majorana communication, regularly published on Nuovo Cimento, and does not rely on more or less arbitrary reconstructions from fragmentary unpublished sources.

It is very surprising that the communication did not receive any mention, neither on the numerous publications of Fermi and his associates on the subject, nor in the further reconstructions of the life and activity of Ettore Majorana.

However, at the end, Fermi eventually fully accepted Majorana improved scheme, and exploited it in the conclusive paper on the subject \cite{FermiAmaldi}.

Therefore, it is clear that Ettore Majorana played a very important role in the development of Fermi statistical model, even without an explicit recognition.

Majorana was deeply involved with research on different topics of atomic, molecular, nuclear, and elementary particle physics, on a time span of at least ten years, from 1928 to 1938. His ideas were very brilliant and his results outstanding, as recognized by all biographers (see in particular the very authoritative account in \cite{Amaldi}, \cite{Amaldi1}). His publication list, even with the addition of the 1928 Rome communication, is rather meager.

Therefore, we started a program of reconstructing on a full scale some of his contributions to different topics, by relying on published sources, and archive sources. This paper gives a first account of his contribution to the improvement of the Thomas-Fermi statistical model for atoms. We plan to report on Majorana contributions to other topics in future publications.
  
\vspace{.5cm}
{\bf Acknowledgments}

We gratefully acknowledge useful conversations with Carlo Bernardini, Luisa Bonolis, Ettore Majorana jr., Bruno Preziosi, and Gilda Senatore, about the human and scientific personality of Ettore Majorana.

We thank the President of the Domus Galilaeana in Pisa, for permission to consult the archives, and to refer to material contained there. 

This work was supported in part by MIUR 
(Italian Minister of Instruction, University and Research), 
and by INFN (Italian National Institute for Nuclear Physics).


\begin{thebibliography}{99}

\bibitem{Majorana} E. Majorana, 
\textit{Ricerca di un'espressione generale delle correzioni di Rydberg, valevole per atomi neutri o ionizzati positivamente}, 
Nuovo Cimento {\bf 6}, XIV-XVI (1929).

\bibitem{Amaldi} E. Amaldi,
\textit{La vita e l'opera di Ettore Majorana},
Accademia Nazionale dei Lincei, Rome, 1966.

\bibitem{Amaldi1} E. Amaldi, 
\textit{Ettore Majorana: Man and Scientist}, pp. 10-75, in: \textit{Strong and weak Interactions. Present problems}, A. Zichichi, ed.,  Academic Press, New York, 1966.

\bibitem{Bonolis} L. Bonolis, \textit{Majorana, il genio scomparso}, Le Scienze S.p.A., Milano, 2002.

\bibitem{Fermi}  E. Fermi, \textit{Un metodo statistico per la determinazione di alcune propriet\`a dell'atomo}, Rend. Accad. Lincei {\bf 6}, 602-607 (1927), (n. 43 in \cite{collected}).

\bibitem{collected} E. Fermi, \textit{Note e Memorie (Collected Papers)}, Vol. I, Italia 1921-1938, Accademia Nazionale dei Lincei, The University of Chicago Press, Roma-Chicago-Londra, 1962. 

\bibitem{Fermiioni} E. Fermi, \textit{Sul calcolo degli spettri degli ioni}, Mem. Accad. Italia, {\bf I} (Fis.), 149-156 (1930), (n. 63 in \cite{collected}).

\bibitem{FermiAmaldi} E. Fermi ed E. Amaldi, \textit{Le orbite $\infty$s degli elementi}, Mem. Accad. Italia, {\bf 6} (Fis.), 119-149 (1934), (n. 82 in \cite{collected}).

\bibitem{AGR} G. Acocella, F. Guerra and N. Robotti, \textit{Enrico Fermi's discovery of neutron-induced artificial radioactivity: the recovery of his first laboratory notebook}, Physics in Perspective {\bf 6}, 29-41 (2004). 

\bibitem{transuranic}  E. Amaldi, O. D'Agostino, E. Fermi, F. Rasetti, E. Segr\`e, \textit{Radioattivit\`a ``beta'' provocata da bombardamento di neutroni. - III}, Ricerca Scientifica {\bf 5}, 452-453 (1934), (n. 86a-86b in \cite{collected}).   

\bibitem{LRS} M. Leone, N. Robotti and C.A. Segnini, \textit{Fermi archives at the Domus Galilaeana in Pisa}, Physis, 501-533 (2000).

\bibitem{Thomas} L.H. Thomas, \textit{The calculation of atomic fields},  Proc. Cambridge Phil. Society, {\bf 23}, 542-548 (1927).
 
\bibitem{Rasetti}  F. Rasetti, \textit{Eine statistische Berechnung der M-R\"ontgenterme}, Zeit. f. Phys. {\bf 48}, 546-549 (1928).

\bibitem{GentileMajorana} G. Gentile and E. Majorana, \textit{Sullo sdoppiamento dei termini Roentgen e ottici a causa dell'elettrone rotante e sulle intensit\`a delle righe del cesio}, Rend. Accad. Lincei {\bf 8}, 229-233 (1928).

\bibitem{FermiFalkenhagen} E. Fermi, \textit{\"Uber die Anwendung der statistischen Methode auf die Probleme des Atombaues}, pp. 95-111, in: \textit{Quantentheorie und Chemie (Leipziger Vortr\"age 1928)}, H. Falkenhagen, ed., S. Hirzel, Leipzig, 1928, (n. 49 in \cite{collected}). 

\bibitem{Gombas} P. Gombas, \textit{Die statistische Theorie des Atoms und ihre Anwendungen}, Springer Verlag, Berlin, 1949.

\bibitem{March} N.H. March, \textit{The Thomas-Fermi approximation in quantum mechanics}, Adv. in Phys. {\bf 6}, 1-98 (1957).

\bibitem{Spruch} L. Spruch, \textit{Pedagogic notes on Thomas-Fermi theory (and on some improvements): atoms, stars and the stability of bulk matter}, Rev. Mod. Physics {\bf 63}, 151-209 (1991).

\bibitem{Miranda} C. Miranda, \textit{Teoremi e metodi per l'integrazione numerica della equazione differenziale di Fermi},
Mem. Accad. Italia {\bf 5}, 285-322 (1934).

\bibitem{Lieb} E. Lieb, \textit{Thomas-Fermi and related theories of atoms and molecules}, Rev. Mod. Physics {\bf 53}, 603-642 (1981).  

\bibitem{Siedentop} H. Siedentop and R. Weikard, \textit{On the leading energy correction for the statistical model of the atom: interacting case}, Commun. Math. Phys. {\bf 103}, 471-490 (1987). 

\bibitem{FermiSegre} E. Fermi e E. Segr\`e, \textit{Sulla teoria delle strutture iperfini}, Mem. Accad. Italia {\bf 4} (Fis.), 131-158 (1933), (n. 75b in \cite{collected}).

\bibitem{Segre} E. Segr\`e, \textit{Calcolo statistico dello spettro di un atomo ionizzato},
Rend. Accad. Lincei {\bf 6}, 670-673 (1930). 

\bibitem{MajoranaKern} E. Majorana, \textit{\"Uber die Kerntheorie}, Zeit. f. Phys. {\bf 82}, 137-145 (1933).  

\bibitem{Heisenberg} W. Heisenberg, \textit{Consid\'erations th\'eoretiques g\'en\'erales sur la structure du noyaux}, pp. 289-324, in: \textit{Structure et propri\'et\'es des noyaux 
            atomiques. Rapports et discussions du VII Conseil de  Physique, Institut Solvay, 22-29 Octobre 1933}, Gauthier-Villars, Paris, 1934. 

\bibitem{BGentile} B. Gentile, \textit{Lettere inedite di E. Majorana a G. Gentile jr.}, Giornale critico della filosofia italiana (Firenze, 1988), p. 145. 

\bibitem{Braunbeck} W. Braunbeck, \textit{Beziehungen der empirischen Atom- und Ionenradien zu der Thomas-Fermischen Ladungsverteilung im Atom}, Zeit. f. Phys. {\bf 79}, 701-710 (1932).

\bibitem{Esposito} S. Esposito, \textit{Majorana solution of the Thomas-Fermi equation},  Am. J. Phys., {\bf 70}, 852-856 (2002).

\end{thebibliography}
\end{document}